\documentclass[11pt]{article}

\usepackage[a4paper,margin=27mm]{geometry}
\usepackage[T1]{fontenc}
\usepackage[utf8]{inputenc}
\usepackage{amsmath}
\usepackage{amsthm}
\usepackage{booktabs}
\usepackage{caption}
\usepackage{enumitem}
\usepackage{float}
\usepackage{graphicx}
\usepackage{microtype}
\usepackage[section]{placeins}
\usepackage{siunitx}
\usepackage{xurl}
\usepackage[hidelinks]{hyperref}

\graphicspath{{../figures/}{figures/}}
\setlist[itemize]{leftmargin=1.35em}
\setlist[enumerate]{leftmargin=1.65em}
\newtheorem{proposition}{Proposition}
\newtheorem{definition}{Definition}

\title{\textbf{Non-Uniform L2 Cache Latency Across the Streaming
Multiprocessors of an NVIDIA L40}}

\author{Faruk Alpay\textsuperscript{*} \quad Barış Başaran\\[0.4em]
\small Department of Computer Engineering, Bahçeşehir University\\
\small Istanbul, Turkey\\
\small \{\texttt{faruk.alpay}, \texttt{baris.basaran}\}\texttt{@bahcesehir.edu.tr}}
\date{}

\begin{document}
\maketitle
\begingroup
\renewcommand{\thefootnote}{\fnsymbol{footnote}}
\footnotetext[1]{Corresponding author: \texttt{alpay@lightcap.ai}}
\endgroup

\begin{abstract}
The NVIDIA L40 exposes a 96\,MiB L2 cache usually modeled as one uniform pool with
a single hit latency. We show this is wrong at the granularity a kernel sees:
L2-hit latency depends strongly and reproducibly on which physical streaming
multiprocessor (SM) issues the load. A turn-serialized, \texttt{\%smid}-resolved
probe maps the hit latency across all 142 SMs in one launch; it is not a constant
near 279 cycles but spans 222--339 cycles (a 52\% range), with per-repetition
noise below \num{0.01} cycles. An additive model
$L=\mu+a(\mathrm{sm})+b(\mathrm{slice})$ explains $R^2=0.87$ (0.98 with one rank-1
term), and the SM term is two-fold symmetric---two halves of 72 SMs at correlation
$r=0.999$---following the AD102 GPC layout. Independent access patterns agree per
SM at $r=1.000$, so the effect is physical. The same probe on a Blackwell RTX 5090
shows it generalizes, while the per-die pattern is device-specific. Read as a
fingerprint, a single user-level probe identifies the SM within a device at 92\%,
and two physically identical L40s are separated at 100\% despite near-identical
mean latency (per-SM map $r=0.63$)---a per-die hardware identity, not a clock
artifact. This is a self-localization and fingerprinting primitive: a kernel reads
its own placement and device, not a victim's, and extracts no secret data. The map
is stable, unchanged after an hour at full utilization on both devices. As a
consequence, distributing latency-bound work by the map cuts makespan by up to
11\%. Single-thread capacity, line-tag, prefetch-modifier, and persisting-L2
results appear as controls. The artifact contains seeds, raw observations, the
trained model, and regeneration scripts.
\end{abstract}

\section{Introduction}

The NVIDIA L40 is a data-center Ada Lovelace GPU built from the full AD102
implementation. CUDA reports a \SI{96}{MiB} L2 cache and a single device-wide
L2; performance models and teaching material routinely treat that cache as a
uniform pool with one access latency. The interface gives no reason to expect
otherwise: there is one \texttt{l2CacheSize}, one persisting set-aside, and one
nominal hit cost.

This abstraction hides a physical fact. The L2 is not a monolith; it is a set of
slices attached to memory controllers, reached from the SMs through an on-chip
network (NoC). Distance through that fabric is not free. Prior measurement work
on real GPUs established that NoC latency between cores and memory partitions is
non-uniform and depends on physical placement, with up to roughly 70\%
variation across V100, A100, and H100 \cite{jin2024noc}. That study did not
cover any Ada GPU, and it emphasized that on those parts the per-core
\emph{average} latency is approximately uniform while the non-uniformity appears
only between specific core--slice pairs, with an additional discrete step when
traffic crosses the A100/H100 dual-partition boundary. Whether a single-die,
single-partition Ada data-center GPU exhibits a comparable structure---and what
form it takes without the partition split---was open.

We answer this for the L40 by measuring L2-hit latency from every physical SM.
The central question is: \emph{is the application-visible L2-hit latency on this
GPU a single number, or a function of physical SM placement, and if so what is
its structure?} The answer is that placement governs the hit latency. We map the
full SM$\times$slice latency space, fit a compact additive model, recover a
two-fold symmetry that matches the AD102 floorplan, confirm the effect against
an independent access pattern, and quantify the resulting timing leakage. The
methodology follows established GPU microbenchmarking
\cite{mei2017dissecting,jia2018volta,luo2025hopper,jin2024noc}; the contribution
is the L40-specific topology and its consequences, including a reinterpretation
of single-thread residency numbers that look like clean uniform-cache results
until placement is taken into account.

The contributions are:

\begin{enumerate}
  \item A per-SM L2-hit latency map of one L40 covering all 142 SMs and 256 slice
        probes, showing a \num{222.5}--\num{339.2}-cycle (52\%) spread where a
        uniform model predicts a constant.
  \item A compact additive model
        $L(\mathrm{sm},\mathrm{slice})=\mu+a(\mathrm{sm})+b(\mathrm{slice})$
        ($R^2=0.87$) that separates an SM-placement term (\num{57.2} cycles)
        from a slice-placement term (\num{39.5} cycles), and a near-low-rank
        refinement (one rank-1 interaction reaches $R^2=0.98$) showing the map is
        a two-coordinate geometry following the GPC layout.
  \item A two-fold symmetry of the SM term: two halves of 72 SMs with profiles
        correlated at $r=0.999$, consistent with the AD102 two-cluster GPC
        layout, together with evidence that the L40 lacks the discrete
        inter-partition latency step of A100/H100 but, unlike those parts, has a
        non-uniform per-SM average.
  \item A robustness argument: single-line slice probes and a \SI{2}{MiB} random
        chain agree at $r=1.000$ per SM, and the symmetry pairs rule out a
        measurement-order confound; per-rep noise is below \num{0.01} cycles.
  \item A side-channel quantification and a trained placement oracle: a single
        L2-latency probe localizes the executing SM to one of roughly 70--120
        classes ($\approx$6--7 bits), and a published classifier identifies the
        exact SM (one of 142) on 99.2\% of held-out fingerprints.
  \item Evidence that the effect generalizes: the same probe on a Blackwell RTX
        5090 finds a 46\% spread and an absolutely slower L2, and shows the
        learned model is device-specific (chance accuracy across devices).
  \item Same-model device fingerprinting: two physically distinct, identical-model
        L40s are separated at 100\% from their per-SM signature despite a
        \num{0.28}-cycle mean-latency difference (per-SM map $r=0.63$), and the
        per-SM oracle does not transfer between dies (0\% vs 98.6\% native)---each
        die has a stable hardware identity.
  \item A user-level self-localization and physical-location primitive: from one
        fingerprint a kernel identifies its own SM within a device at 92\% and its
        device, using no privileges or counters. It is self-localization, not
        victim or co-resident inference, and extracts no secret data.
  \item A stability result: the per-SM map is unchanged after an hour at 100\%
        utilization on both devices (correlation \num{1.000}, drift below
        \num{0.4} cycles), so the topology is a property of the silicon.
  \item A measured consequence: scheduling latency-bound work by the per-SM map
        reduces makespan by up to 11\%, with no benefit in the DRAM-bound regime.
  \item Single-thread residency controls---capacity boundary, line-tag scaling,
        a PTX prefetch-modifier null result, and a persisting-L2 boundary---now
        read as averages over the non-uniform latency they each assume to be a
        constant.
\end{enumerate}

\section{Method: Per-SM L2 Latency Probing}
\label{sec:method}

Table~\ref{tab:system} summarizes the tested instance. The GPU was accessed
through SSH on a Vast.ai CUDA 13.0 base image. The container runs without
elevated privileges, so
hardware performance counters were not available through Nsight Compute on this
host \cite{nsight-profiling-guide}; the artifact relies on timing, CUDA device
properties, \texttt{nvidia-smi} telemetry, source hashes, command lines, and
repeated raw observations.

\begin{table}[H]
\centering
\caption{System under test. Values are recorded in the run manifests.}
\label{tab:system}
\begin{tabular}{ll}
\toprule
Component & Observed property \\
\midrule
GPU & NVIDIA L40, compute capability 8.9 \\
SMs & 142 \\
Global memory & \SI{47665709056}{B} visible (\SI{44.39}{GiB}) \\
L2 cache & \SI{100663296}{B} (\SI{96}{MiB}) \\
Persisting L2 maximum & \SI{69206016}{B} (\SI{66}{MiB}) \\
Access-policy max window & \SI{134213632}{B} (\SI{127.996}{MiB}) \\
Shared memory per SM & \SI{102400}{B} (\SI{100}{KiB}) \\
Memory interface & \SI{384}{bit} GDDR6, \SI{9001}{MHz} \\
Reported SM clock & \SI{2490}{MHz} \\
ECC & Enabled \\
Driver / runtime & \num{580.142} / CUDA 13.0 \\
Host CPU & Intel Xeon Gold 5418Y, 48 visible cores \\
\bottomrule
\end{tabular}
\end{table}

\paragraph{Probe.} Each measurement times a single-thread, single-warp dependent
load chain, following the standard one-request-in-flight design that removes
coalescing and queueing so the timed value is one round trip through the cache
fabric \cite{jin2024noc,mei2017dissecting}. The load is \texttt{ld.global.cg},
which uses the global cache path and bypasses the private L1, so a hit is an L2
hit. Two buffer shapes are used. The \emph{single-line} shape fills a buffer with
$\mathtt{buf}[i]=i$, so any start word loops forever on its own \SI{128}{B} line;
repeated loads of that one line are L2 hits to the single slice that owns the
address, and sweeping the byte offset moves the probe across slices. The
\emph{chain} shape is a random permutation pointer chain over a \SI{2}{MiB}
footprint that touches the whole slice ensemble, so its per-SM latency is the
average distance to all slices. For footprint $F$ and stride $S$ in the residency
controls (Section~\ref{sec:residency}), the number of distinct \SI{128}{B} line
tags touched is $T(F,S)=F/128$ for $S<128$ and $F/S$ otherwise, with
tag-equivalent footprint $F_{\mathrm{tag}}=128\,T=F\min(1,128/S)$.

\paragraph{SM attribution and isolation.} Each block reads its physical SM
through the \texttt{\%smid} special register and writes its measurement indexed
by that identifier. One block is forced resident per SM by requesting more than
half of the \SI{100}{KiB} shared memory per SM as dynamic shared memory; the
occupancy calculator then reports one block per SM, so a launch of 142 blocks
covers every SM exactly once. To remove on-chip contention during timing, the
blocks are serialized by a global turn counter: a block claims a turn with
\texttt{atomicAdd}, waits with \texttt{\_\_nanosleep} back-off until its turn is
served, performs warm-up and then times its dependent loads alone, and releases
the turn. Only one block touches memory in its timed region, so the measured
cycles isolate that SM's path. The reported latency is

\[
L(\mathrm{sm},\mathrm{slice})=\frac{\text{end cycles}-\text{begin cycles}}{A},
\]

with $A$ timed loads per block; the slice campaign uses $A=8192$, four
repetitions, and 256 consecutive \SI{128}{B} probes; the chain control uses
$A=8192$ and 16 repetitions.

\paragraph{Model.} We summarize the map with an additive decomposition, in the
spirit of non-uniform cache access (NUCA) models for wire-delay-dominated caches
\cite{kim2002nuca}.

\begin{definition}[Additive NUCA model]
For measured latencies $L(\mathrm{sm},\mathrm{slice})$ with grand mean $\mu$,
let $a(\mathrm{sm})=\overline{L}(\mathrm{sm},\cdot)-\mu$ and
$b(\mathrm{slice})=\overline{L}(\cdot,\mathrm{slice})-\mu$. The additive model is
$\widehat{L}=\mu+a(\mathrm{sm})+b(\mathrm{slice})$, and the residual
$L-\widehat{L}$ is the SM$\times$slice interaction.
\end{definition}

The fraction of variation explained, the ranges of $a$ and $b$, and the residual
standard deviation are reported in Section~\ref{sec:topology}; the interaction is
the part of the latency that depends jointly on which SM reaches which slice.

\section{The L2 Latency Topology}
\label{sec:topology}

\begin{figure}[H]
\centering
\includegraphics[width=\linewidth]{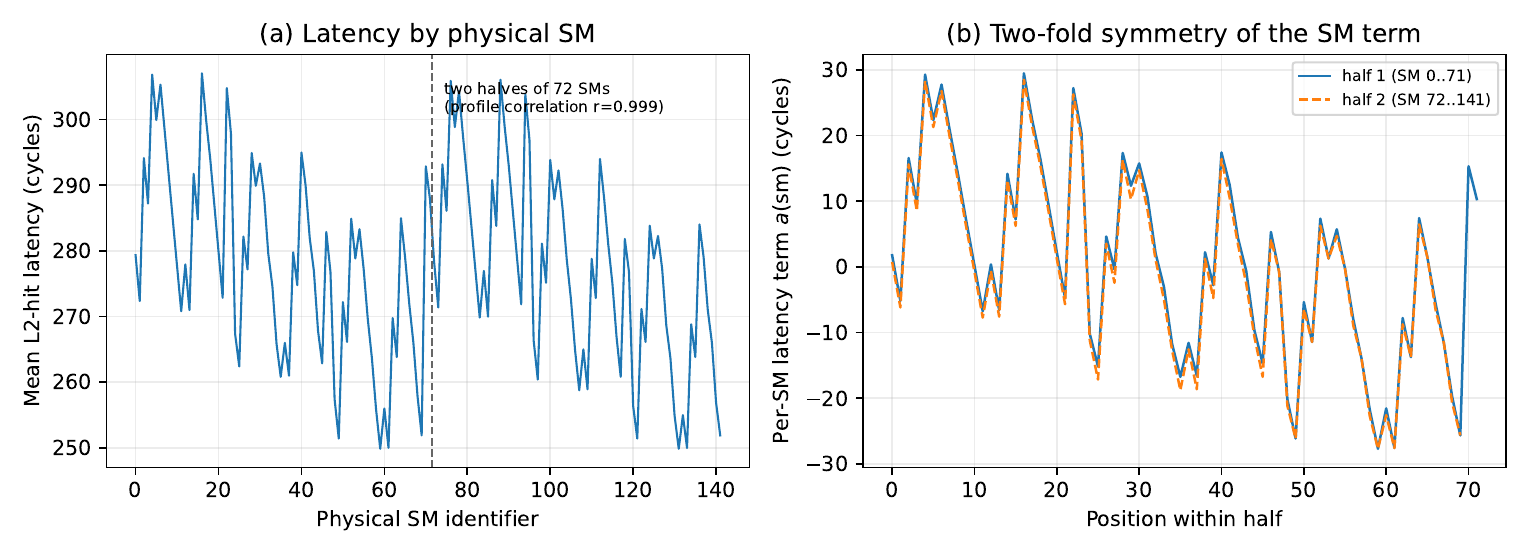}
\caption{Per-SM L2-hit latency on one L40. (a) Mean latency against physical SM
identifier; the dashed line marks the split into two halves of 72 SMs. (b) The
SM-placement term $a(\mathrm{sm})$ for the two halves overlaid; the profiles
agree at correlation $r=0.999$, a two-fold symmetry consistent with the AD102
two-cluster GPC layout.}
\label{fig:persm}
\end{figure}

Figure~\ref{fig:persm} maps the mean L2-hit latency of every physical SM. Panel
(a) plots latency against the SM identifier. The value is not constant: it ranges
from \num{249.8} to \num{307.0} cycles when averaged over slices, and the curve
has visible internal structure rather than noise. Panel (b) shows that this
structure is two-fold symmetric. Splitting the SM-placement term $a(\mathrm{sm})$
at identifier 72 yields two halves of 72 and 70 SMs whose profiles are nearly
identical, with correlation $r=0.999$ and a mean absolute difference of
\num{0.99} cycles; the six fastest SMs occur in exact identifier pairs
$(49,121)$, $(59,131)$, $(61,133)$ separated by 72, and the slowest pair as
$(4,76)$ and $(16,88)$. Within each half the latency varies with position
(correlation $-0.53$ with index), and across the chip the per-SM averages form a
continuous distribution with no bimodal gap. The period of 72 matches six of the
twelve AD102 GPCs at twelve SMs each, so the most direct reading is that the
\texttt{\%smid} enumeration walks two symmetric six-GPC clusters and that an SM's
latency is set by its position within a cluster.

\begin{figure}[H]
\centering
\includegraphics[width=0.92\linewidth]{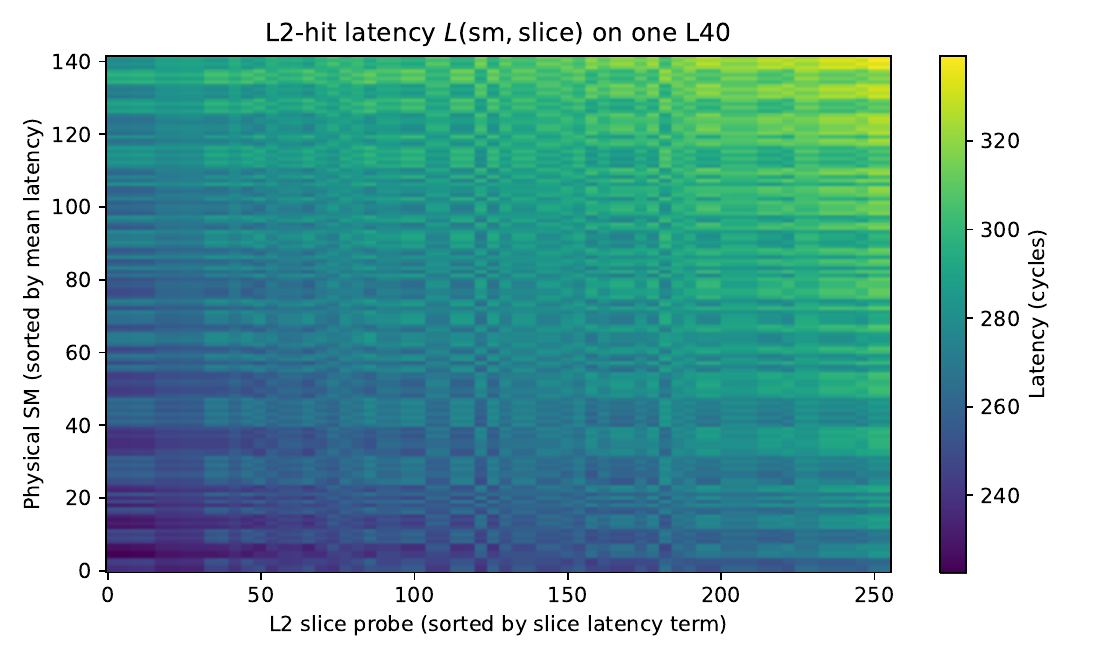}
\caption{L2-hit latency $L(\mathrm{sm},\mathrm{slice})$ over all 142 SMs and 256
slice probes. SMs are ordered by mean latency and probes by slice term. Latency
rises smoothly from \num{222.5} to \num{339.2} cycles; the vertical striping is
the per-slice term with a \SI{512}{B} interleave period.}
\label{fig:heatmap}
\end{figure}

Figure~\ref{fig:heatmap} shows the full latency map over the SM$\times$slice
space, with SMs ordered by mean latency and slice probes ordered by slice term.
The surface is smooth in both directions: latency rises from \num{222.5} cycles
at the near SM / near slice corner to \num{339.2} cycles at the far corner, a
\num{116.6}-cycle (52\%) total range, equivalently \SI{89.4}{ns} to
\SI{136.2}{ns} at the reported \SI{2.49}{GHz}. The vertical striping is the slice
structure: consecutive \SI{128}{B} probes alternate among slices, and the
slice-term autocorrelation has its first strong period at four lines
(\SI{512}{B}), the granularity at which the L2 interleaves addresses across
slices on this part. The two visible axes are exactly the two additive terms; the
fine texture on top of them is the interaction.

\begin{figure}[H]
\centering
\includegraphics[width=\linewidth]{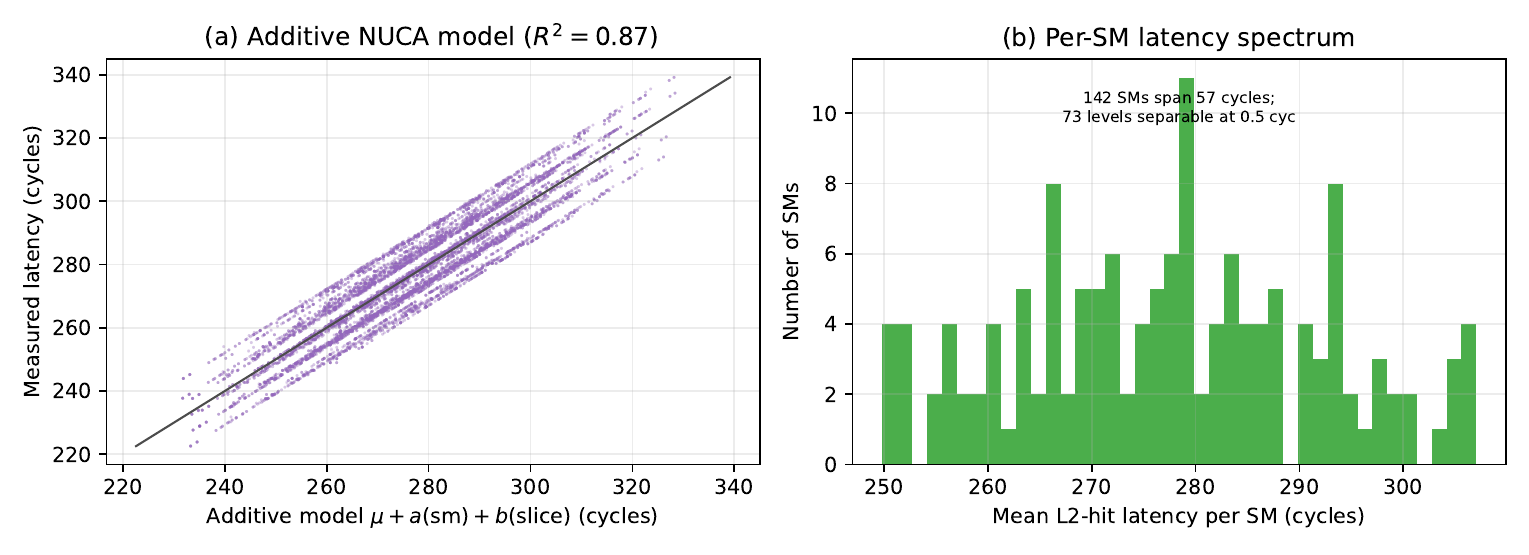}
\caption{(a) Additive model $\mu+a(\mathrm{sm})+b(\mathrm{slice})$ against
measured latency ($R^2=0.87$); the diagonal banding is the SM$\times$slice
interaction. (b) Per-SM latency spectrum: a continuous 57-cycle spread with no
discrete partition step.}
\label{fig:model}
\end{figure}

Figure~\ref{fig:model} quantifies the decomposition. Panel (a) plots measured
latency against the additive prediction $\mu+a+b$; the model explains $R^2=0.87$
of the variation, with an SM term spanning \num{57.2} cycles, a slice term
spanning \num{39.5} cycles, and a residual standard deviation of \num{6.85}
cycles for the interaction. Panel (b) shows the per-SM latency spectrum: the 142
SMs spread continuously over 57 cycles with no partition gap. This is the first
point of contrast with A100/H100, where the dominant non-uniformity is a discrete
step at the partition boundary and the per-core average is approximately uniform
\cite{jin2024noc}. On the single-partition L40 there is no such step, but the
per-SM average is itself non-uniform by 23\% (the chain control independently
measures a \num{57.1}-cycle per-SM spread), so placement matters here even for an
access pattern that touches the whole cache evenly.

Two finer structures sharpen this picture. First, the map is nearly low-rank.
Adding a single rank-1 interaction term $c\,u(\mathrm{sm})\,v(\mathrm{slice})$ to
the additive model raises $R^2$ from 0.87 to 0.98 on the L40 (and from 0.83 to
0.99 on the RTX 5090 of Section~\ref{sec:crossdev}). The interaction's SM
coordinate $u$ is essentially uncorrelated with the additive term $a$
($|r|=0.06$), so it is a second, independent placement axis rather than a
rescaling of the first; two coordinates per SM and per slice thus reproduce the
latency map almost completely. Second, the SM term is hierarchical: its
autocorrelation peaks at a period of 12 SMs on the L40 (and 10 on the RTX 5090),
matching the per-GPC SM counts of the two floorplans, nested inside the two-fold
macro symmetry of Figure~\ref{fig:persm}. The non-uniformity is therefore not
noise on a flat cache but a structured, low-dimensional geometry that follows the
GPC layout.

\paragraph{The effect is physical, not procedural.} Three checks rule out
artifacts. First, per-repetition noise is small: the median standard deviation of
the per-access latency across repetitions is \num{0.006} cycles, so a difference
of tens of cycles between SMs is four orders of magnitude above the noise floor.
Second, the two-fold symmetry decouples placement from measurement order: SMs $k$
and $k+72$ are timed roughly 72 turns apart yet differ by \num{0.99} cycles on
average, whereas an order- or temperature-driven drift would scale with turn
separation. Third, the single-line slice probe and the \SI{2}{MiB} random chain
are different access patterns measured in independent launches with independent
turn orders, and their per-SM averages correlate at $r=1.000$; within a fixed SM,
latency is uncorrelated with turn across repetitions (mean correlation $-0.13$).
The per-SM latency is therefore a stable function of the physical SM.

\section{Timing Leakage of Physical Placement}
\label{sec:sidechannel}

Because the per-SM latency is stable and the measurement noise is small, the
latency itself identifies the SM. We state this as a separability bound.

\begin{proposition}[SM separability]
Let the per-SM mean latencies $\{\overline{L}(\mathrm{sm})\}$ have single-probe
standard deviation $\sigma$. The number of SM classes distinguishable at
confidence level $k\sigma$ is the largest $C$ such that the sorted per-SM means
admit $C-1$ consecutive gaps exceeding $k\sigma$. With $\sigma\le\num{0.01}$
cycles and the measured spread of \num{57.2} cycles, $C\ge 118$ at $k=5$.
\end{proposition}

\noindent The bound follows directly from ordering the per-SM means and counting
gaps wider than $k\sigma$; with the measured $\sigma$ almost every pair of SMs is
separable. A coarser, more conservative count that bins the means at
\num{0.5}-cycle resolution still yields 73 distinct levels. Either way, a single
L2-latency probe localizes the executing SM to one of roughly 70--120 classes,
about 6--7 bits of its \emph{own} physical placement. This is the L40 analogue of
the core-location timing channel that motivates real-GPU NoC analysis
\cite{jin2024noc} and connects to practical GPU side channels
\cite{naghibijouybari2018rendered}: a user-level kernel can infer where it is
physically scheduled, and a scheduler that wanted to hide placement would have to
defeat a 57-cycle signal that sits far above the noise. What is measured is
\emph{self}-localization---a kernel reading its own
placement; inferring the placement of a separate, co-resident victim kernel would
require a contention or interference channel that we do not measure, and no
secret data are recovered. This is a localization primitive---a stepping stone
toward de-anonymization or co-residency detection \cite{dutta2022spy}, not a
complete exploit. The same signal is usable constructively, as a cheap placement
oracle for locality-aware scheduling.

\subsection{A trained placement oracle}

To make the bound concrete and reusable, we train a model that predicts the
physical SM from a measured fingerprint. The fingerprint is a vector of L2-hit
latencies to 32 fixed \SI{128}{B} lines spaced \SI{256}{B} apart, each timed
with $A$ dependent loads; a single launch of the serialized probe yields one
fingerprint per SM, so a ``shot'' is one fingerprint for each of the 142 SMs. We
collect 500 shots at each of $A\in\{32,64,128,256\}$ loads, label each
fingerprint with its \texttt{\%smid}, and split train/test by shot so the test
fingerprints come from shots never seen in training. A random-forest classifier
is trained per condition; the published model uses $A=256$ and all 32 probes.

With a \SI{256}{load} fingerprint the oracle identifies the exact SM, one of 142,
on 99.2\% of held-out shots, with the correct SM in its top five every time and
the two-cluster label recovered at the same 99.2\% against a 0.7\% chance rate;
even a fast \SI{32}{load} fingerprint reaches 96.3\%. A nearest-centroid baseline
already reaches 98.9\%, so the signal, not the model, carries the leakage. The
leakage does not need a long fingerprint either: a single probe localizes the SM
to 75.6\% exact and 90.6\% cluster accuracy, rising to 99.2\% with all 32 probes.
The remaining errors are almost entirely the symmetric twins: on the
held-out sample the only misclassification maps an SM to its $k\!\leftrightarrow\!
k+72$ partner, which Section~\ref{sec:topology} showed are near-identical. The
oracle therefore fails exactly where the fabric is symmetric, and its error mode
re-derives the two-fold structure. We publish the trained model, its probe-bank
metadata, and a held-out sample so the oracle runs offline without a GPU. The
oracle performs \emph{self}-localization: it reports the physical SM of the kernel
that took the fingerprint. It is not an attack on a separate victim---reading
another kernel's placement would need a contention channel we do not measure---and
it extracts no secret data. The same primitive is usable constructively, as a
cheap placement estimate for locality-aware scheduling
(Section~\ref{sec:placement}).

\section{Cross-Architecture Generalization}
\label{sec:crossdev}

To test whether the non-uniformity is specific to the L40, we ran the identical
probe on an NVIDIA RTX 5090 (Blackwell GB202, 170 SMs, the same nominal
\SI{96}{MiB} L2). Table~\ref{tab:crossdev} compares the two devices. The
phenomenon generalizes: the 5090 L2-hit latency also depends on physical SM
placement, spanning 288.2--419.5 cycles (46\%), with an additive model at
$R^2=0.83$ and the same \SI{512}{B} slice-interleave period. Two architectural
differences stand out. First, the 5090 L2 is markedly slower in absolute
time---\SI{119.7}{ns}--\SI{174.3}{ns} versus the L40's
\SI{89.4}{ns}--\SI{136.2}{ns}---despite the identical nominal capacity, which is
consistent with the larger die and higher SM count lengthening the average
SM-to-slice path. Second, the L40's two-fold symmetry is near-perfect
($r=0.999$) while the 5090's is weaker ($r=0.80$ at a split of 88 SMs),
indicating a less symmetric arrangement of SMs around the fabric on GB202.

\begin{table}[H]
\centering
\caption{The per-SM L2 latency topology on two architectures, measured with the
identical probe. The SM/slice terms are the additive-model term ranges in cycles.}
\label{tab:crossdev}
\small
\setlength{\tabcolsep}{5pt}
\begin{tabular}{lrrrrrr}
\toprule
Device & SMs & L2-hit (ns) & Spread & $R^2$ & SM/slice (cyc) & 2-fold $r$ (split) \\
\midrule
L40 (Ada AD102)            & 142 &  89.4--136.2 & 52\% & 0.87 & 57 / 40 & 0.999 (72) \\
RTX 5090 (Blackwell GB202) & 170 & 119.7--174.3 & 46\% & 0.83 & 64 / 46 & 0.80 (88) \\
\bottomrule
\end{tabular}
\end{table}

\begin{figure}[H]
\centering
\includegraphics[width=\linewidth]{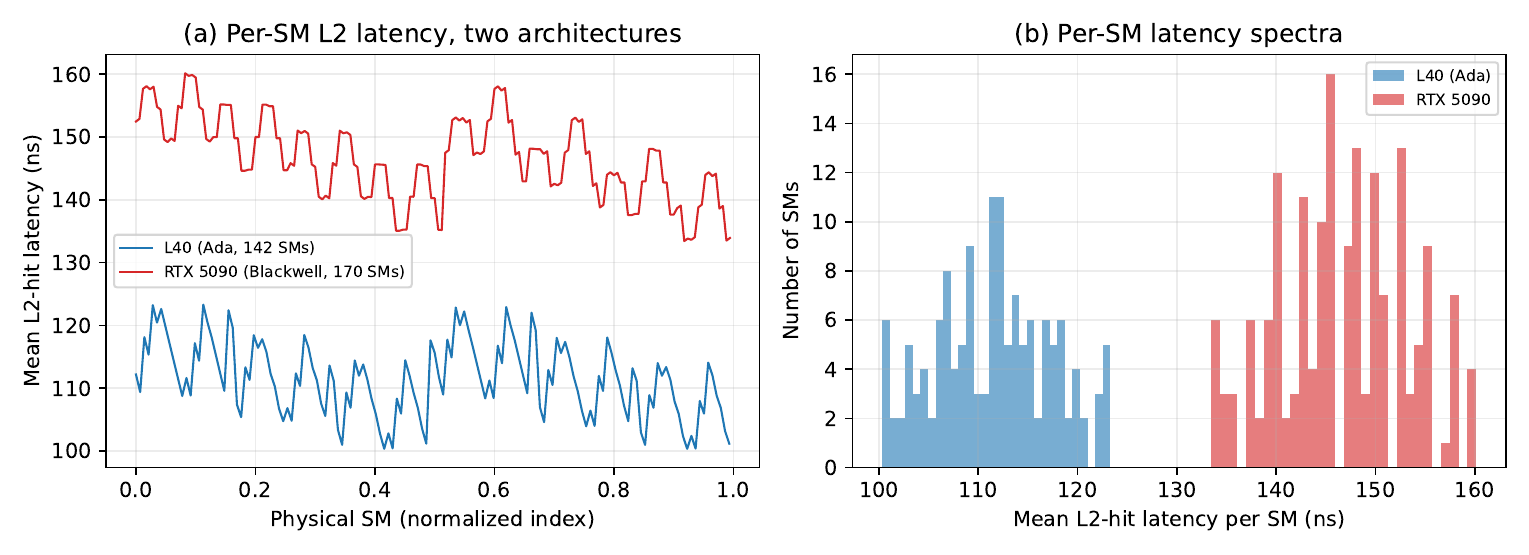}
\caption{Per-SM L2-hit latency on the Ada L40 and the Blackwell RTX 5090,
measured with the same probe. (a) Latency by physical SM (normalized index);
both are non-uniform, and the 5090 sits higher. (b) Per-SM latency spectra in
nanoseconds; the two devices occupy disjoint bands.}
\label{fig:crossdev}
\end{figure}

The fingerprint is therefore architecture-specific. The L40-trained oracle of
Section~\ref{sec:sidechannel}, applied to 5090 fingerprints, drops to 0.6\%
accuracy---exactly chance for 170 classes---while a 5090-native oracle trained by
the same procedure reaches 86.2\%. The method transfers across architectures; the
learned model does not. This confirms that the latency map is a genuine device
signature and that the central finding is not an L40 artifact.

\section{Device Fingerprinting and Physical-Location Inference}
\label{sec:locator}

Because the per-SM map is specific to each physical device, it acts as a
fingerprint. We separate two questions: which \emph{device} a kernel runs on, and
where on it (which SM).

\subsection{Same-model device fingerprinting}
\label{sec:samemodel}

\begin{figure}[H]
\centering
\includegraphics[width=0.62\linewidth]{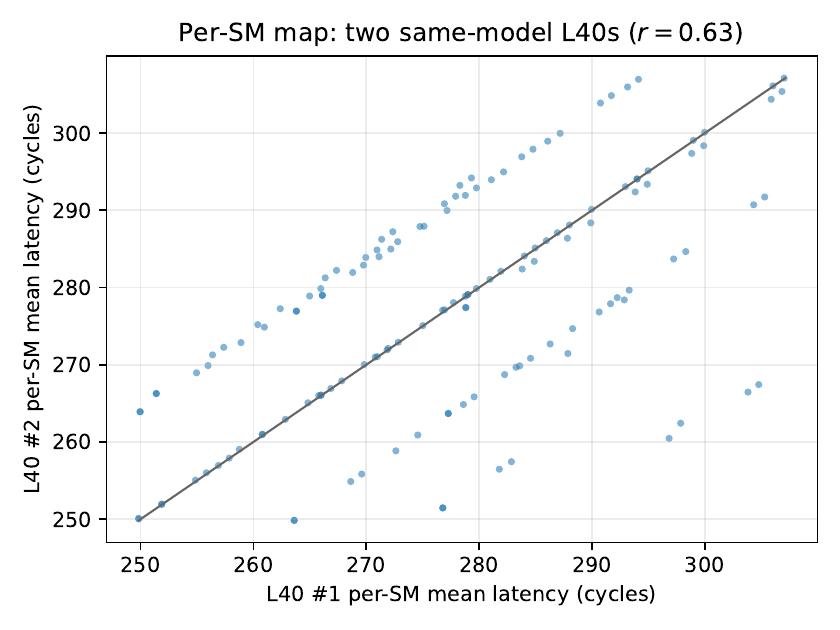}
\caption{Two physically distinct, identical-model L40s: their per-SM latency maps
correlate at only $r=0.63$ despite near-identical means, so each die has its own
pattern. The two devices are separated at 100\% from these fingerprints and the
per-SM oracle does not transfer between dies (figures in text).}
\label{fig:samemodel}
\end{figure}

The cross-architecture case is easy: the L40 and RTX 5090 latency bands are
disjoint (Section~\ref{sec:crossdev}), so a fingerprint names the architecture
trivially. The substantive question is whether two \emph{identical-model} devices
are distinguishable. We measured a second physical L40---distinct UUID, different
driver---with the same probe. The two L40s have nearly identical global mean
latency (\num{277.5} versus \num{277.8} cycles, a \num{0.28}-cycle offset), so
they cannot be told apart by average latency or clock. Yet their per-SM maps
differ markedly: the maps correlate at only $r=0.63$
(Figure~\ref{fig:samemodel}), and after removing the global offset the per-SM
difference has a standard deviation of \num{12.4} cycles (up to \num{37.7}), far
above the \num{0.006}-cycle noise floor. A classifier separates the two devices
at 100\%, and stays at 100\% after de-meaning each fingerprint, so the signal is a
genuine per-die pattern rather than a global offset. The placement oracle is
itself die-specific: the first-L40 oracle scores 0\% on the second (below the
0.7\% chance rate), while a second-L40-native oracle reaches 98.6\%. Each physical
L40 thus carries a stable, distinct
per-SM L2 signature---a hardware identity---most plausibly from the combination of
process variation and the device's particular disabled-SM pattern (each L40
enables 142 of the 144 AD102 SMs).

\subsection{Physical-location inference}

\begin{figure}[H]
\centering
\includegraphics[width=0.62\linewidth]{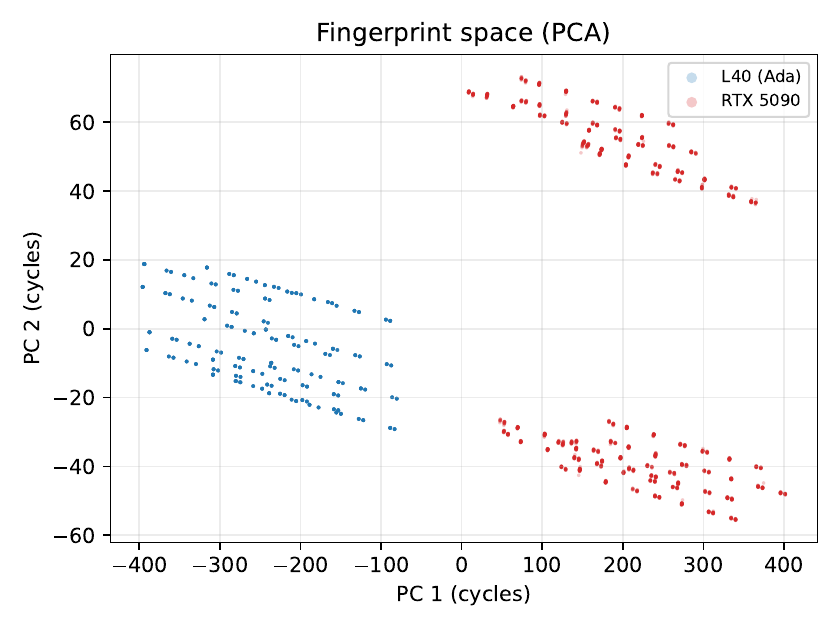}
\caption{Physical-location inference: the fingerprint space (first two principal
components); the two devices occupy disjoint regions, each with an internal per-SM
gradient. The 312-way location accuracy (64.6\% from one probe, 92.1\% from 32) is
given in the text.}
\label{fig:locator}
\end{figure}

Pooling the first L40 and the RTX 5090 into a 312-location label space (142 + 170
SMs), one classifier on the shared 32-probe feature recovers the full physical
location on 92.1\% of held-out fingerprints against a 0.3\% chance rate; a single
probe already reaches 64.6\%. Here the device split is the easy, cross-architecture
case---the two devices occupy disjoint regions of fingerprint space
(Figure~\ref{fig:locator})---and the substantive part is the within-device SM,
identified at 92.1\%. No privileges, performance counters, or topology APIs are
used---the only input is the timing of ordinary global loads.

\paragraph{Scope.} What is demonstrated is
\emph{self}-localization and device identification: a kernel reads its own
physical placement and the identity of the device it runs on. This is a
localization and fingerprinting \emph{primitive}---a stepping stone toward
de-anonymization or co-residency detection \cite{dutta2022spy}, not a complete
exploit. We do not infer the placement of a separate victim kernel: that would
require a contention or interference channel that we do not measure, and we
extract no secret data. The same primitive is the security face of the property
that drives the performance result of Section~\ref{sec:placement}.

\section{A Consequence: NUCA-Aware Work Placement}
\label{sec:placement}

If L2-hit latency depends on SM placement, then distributing latency-bound work
uniformly wastes time on the slow SMs---the GPU analogue of locality-aware data
placement in distributed CPU caches \cite{hardavellas2009rnuca}. We test this
directly. A fixed amount of
dependent-load work is spread across the SMs under three policies, and the
wall-clock makespan is measured with CUDA events: \emph{oblivious} (equal work
per SM, no topology knowledge), \emph{aware} (work per SM proportional to
$1/a(\mathrm{sm})$ from the measured map), and \emph{dynamic} (a global atomic
work queue that balances itself at runtime without a model). The total work is
identical across policies; only its distribution differs.

\begin{figure}[H]
\centering
\includegraphics[width=0.78\linewidth]{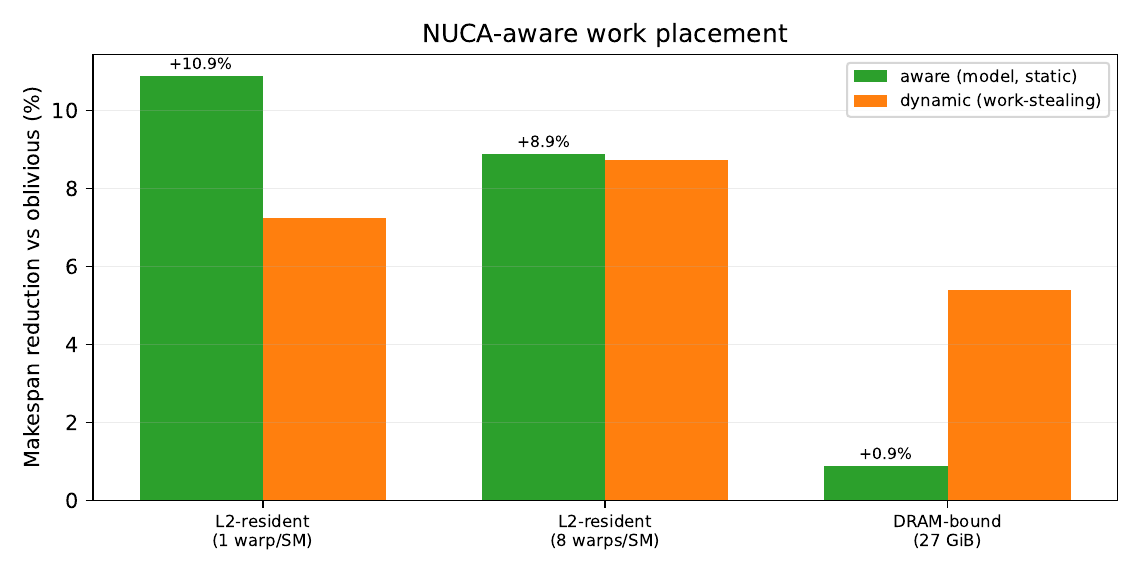}
\caption{Makespan reduction over the oblivious baseline for a fixed latency-bound
workload. The model-based static schedule (aware) matches or beats runtime
work-stealing in the L2-resident regime and gives essentially nothing once the
workload is DRAM-bandwidth bound.}
\label{fig:placement}
\end{figure}

Figure~\ref{fig:placement} reports the result. When the working set is
L2-resident and the workload is latency-bound, the aware schedule reduces
makespan by 10.9\% with one warp per SM and 8.9\% with eight, matching the
$\max_i a_i / \mathrm{HM}(a)$ ratio predicted by the per-SM map. The model-based
static schedule equals or slightly beats the dynamic work-stealing scheduler
(7.3--8.7\%) while incurring no runtime coordination. The gain is bounded to the
regime where it should exist: when the per-thread footprint is enlarged to
\SI{27}{GiB} so the workload is DRAM-bandwidth bound, the aware schedule gives
only 0.9\%, because L2 latency no longer governs throughput. The non-uniform
topology thus has a regime-specific scheduling consequence: it matters for
latency-bound L2 reuse and not for streaming.

\section{Stability Under Sustained Load}
\label{sec:stability}

A fingerprint or a scheduling decision is only useful if the topology is stable.
We held each device at full utilization for one hour---a \SI{38}{GiB} buffer on
the L40 and a \SI{26}{GiB} buffer on the RTX 5090, hammered by a compute-and-memory
load kernel---and took a 32-probe per-SM snapshot every \SI{60}{s}, 60 snapshots
per device.

\begin{figure}[H]
\centering
\includegraphics[width=0.62\linewidth]{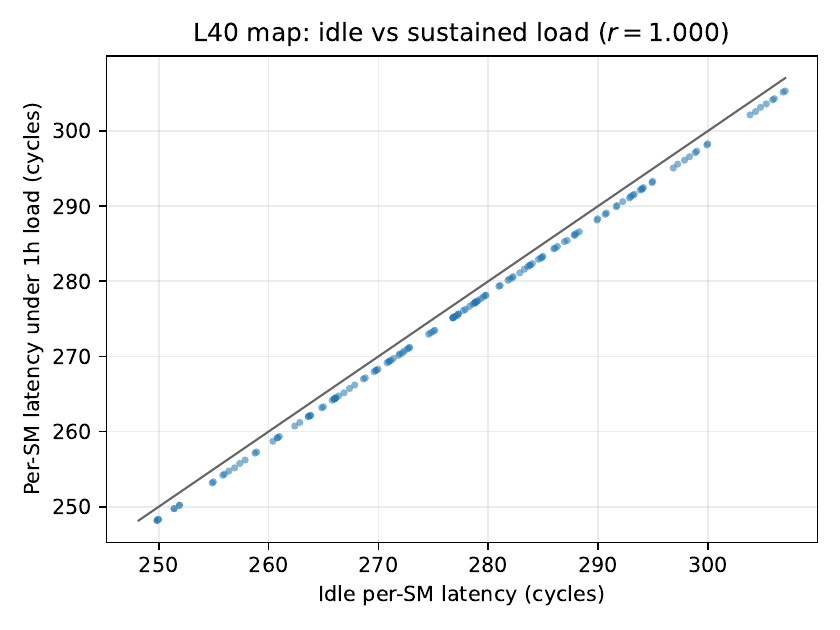}
\caption{The L40 per-SM map after one hour of full GPU and VRAM load matches the
independently measured idle map at $r=1.000$. Across the run the
snapshot-to-snapshot correlation stays at 1.000 on both devices (in text).}
\label{fig:stability}
\end{figure}

The per-SM map is essentially invariant. On the L40 the snapshot-to-snapshot
correlation has median \num{1.000} and the per-SM mean drifts by at most
\num{0.08} cycles over the hour at 100\% utilization and \SI{63}{\celsius}; the
loaded map matches the independently measured idle map at $r=1.000$
(Figure~\ref{fig:stability}). The RTX 5090 behaves the same (correlation
\num{1.000}, drift \num{0.35} cycles, 100\% utilization, \SI{64}{\celsius},
\SI{402}{W}, \SI{2857}{MHz}). The topology is therefore a property of the silicon,
not of a transient idle state. One nuance bounds the fingerprint's portability:
an oracle trained on idle fingerprints transfers poorly to under-load fingerprints
(8.5\% on the L40), because the fine per-probe detail shifts with the operating
condition even though the per-SM mean does not; a load-calibrated oracle recovers
91.4\%. The signature persists under load, but a deployed locator should be
calibrated to the operating point.

\section{Single-Thread Residency Controls}
\label{sec:residency}

The remaining experiments are single-thread residency measurements. Each one
implicitly treats the L2-hit latency as one constant near 279 cycles; the
topology result shows that this constant is the average of a
\num{222.5}--\num{339.2}-cycle distribution and, for a one-block probe, the value
of whichever SM the scheduler happened to use. The residency conclusions survive
this reinterpretation because they depend on the \emph{transition} between the
L2-hit and DRAM regimes, not on the absolute hit latency, but the framing
matters: a ``single hit latency'' is a per-SM quantity.

\begin{figure}[H]
\centering
\includegraphics[width=0.82\linewidth]{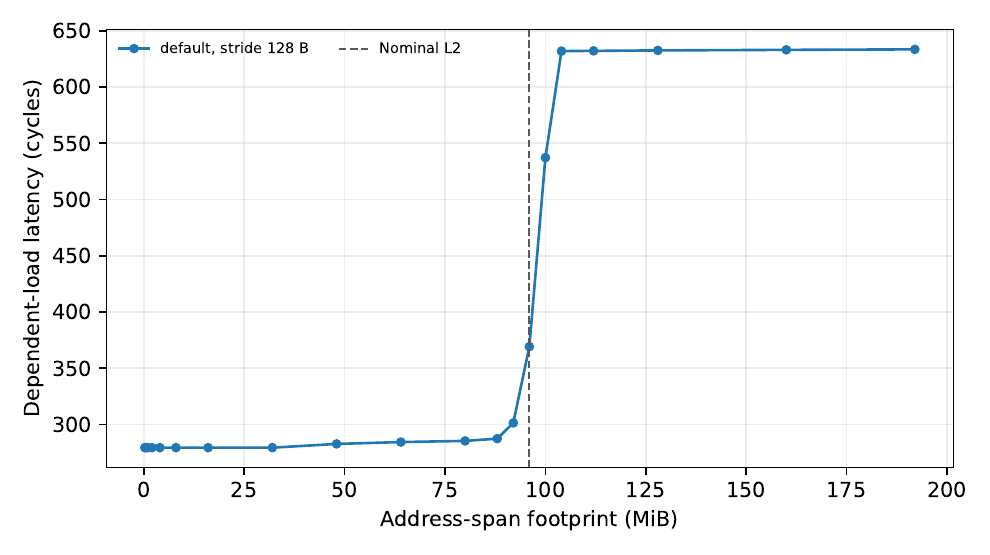}
\caption{Dependent-load capacity sweep (\SI{128}{B} stride). The median rises
from \num{301.3} cycles at \SI{92}{MiB} to \num{631.9} cycles at \SI{104}{MiB},
bracketing the nominal \SI{96}{MiB} L2 boundary.}
\label{fig:capacity}
\end{figure}

Figure~\ref{fig:capacity} shows a capacity sweep with a \SI{128}{B} stride. The
median latency is flat through \SI{32}{MiB} at \num{279.3} cycles, crosses a
narrow transition near the nominal \SI{96}{MiB} boundary (\num{301.3} cycles at
\SI{92}{MiB}, \num{369.1} at \SI{96}{MiB}, \num{537.2} at \SI{100}{MiB}), and
plateaus near 633 cycles in the DRAM-backed regime. This establishes the two
regimes; it does not by itself identify the replacement variable, because a dense
chain changes address span and unique tag count together.

Varying the stride separates the two. Table~\ref{tab:tagmidpoints} lists the
interpolated transition midpoints. Raw address-span midpoints cover a
\num{7.6}$\times$ range (\num{98.16} to \SI{745.48}{MiB}), but normalizing to
\SI{128}{B} tag-equivalent footprint collapses them to
$\num{98.56}\pm\SI{3.46}{MiB}$ (coefficient of variation 3.5\%), near the
physical \SI{96}{MiB} capacity. Replacement is governed by unique \SI{128}{B}
line tags, not by virtual address span---a consistency check against the
line-tag model rather than a new result.

\begin{table}[H]
\centering
\caption{Capacity-transition midpoints for different strides.}
\label{tab:tagmidpoints}
\begin{tabular}{rrrr}
\toprule
Stride (B) & Raw midpoint (MiB) & Tag-equivalent midpoint (MiB) & Midpoint cycles \\
\midrule
32   & 102.46 & 102.46 & 457.99 \\
64   & 103.05 & 103.05 & 454.98 \\
128  &  98.16 &  98.16 & 459.18 \\
256  & 197.26 &  98.63 & 460.97 \\
512  & 383.49 &  95.87 & 460.08 \\
1024 & 745.48 &  93.18 & 443.80 \\
\bottomrule
\end{tabular}
\end{table}

PTX accepts load modifiers that request L2 prefetch sizes
\cite{ptx-isa}. Compiling the same dependent-chain kernel with default
\texttt{ld.global.cg} and with \texttt{L2::64B}, \texttt{L2::128B}, and
\texttt{L2::256B} does not move the replacement boundary:
Table~\ref{tab:prefetch} shows all four forms crossing at \SI{98.1}{MiB} for
\SI{128}{B} stride and \SI{206.0}{MiB} for \SI{256}{B} stride, differing only by
a few cycles in the high plateau. The governing capacity variable remains the
unique line-tag count.

\begin{table}[H]
\centering
\caption{Midpoint crossings for prefetch modifiers.}
\label{tab:prefetch}
\begin{tabular}{llrr}
\toprule
Load form & Stride & Midpoint (MiB) & High plateau (cycles) \\
\midrule
default & \SI{128}{B} & 98.09 & 634.95 \\
\texttt{L2::64B} & \SI{128}{B} & 98.14 & 637.23 \\
\texttt{L2::128B} & \SI{128}{B} & 98.14 & 641.66 \\
\texttt{L2::256B} & \SI{128}{B} & 98.12 & 641.69 \\
default & \SI{256}{B} & 206.02 & 635.07 \\
\texttt{L2::64B} & \SI{256}{B} & 206.02 & 637.21 \\
\texttt{L2::128B} & \SI{256}{B} & 206.03 & 641.67 \\
\texttt{L2::256B} & \SI{256}{B} & 206.04 & 641.66 \\
\bottomrule
\end{tabular}
\end{table}

Finally, the CUDA persisting-L2 set-aside has a measurable boundary. CUDA reports
a \SI{66}{MiB} maximum set-aside on this L40 \cite{cuda-l2-control}. With a
\SI{256}{MiB} cold stream between warm-up and probe, Table~\ref{tab:retention}
shows that persisting policy keeps 16--64 MiB hot sets in the L2-hit regime,
partially protects \SI{72}{MiB}, and gives no benefit for 80--88 MiB hot sets,
matching the reported set-aside. This is a measured sufficient condition for a
serialized hot/cold pattern, not a universal replacement guarantee.

\begin{table}[H]
\centering
\caption{Persisting-L2 hot-set boundary after a \SI{256}{MiB} cold stream
(\SI{128}{B} stride, $10^6$ probe loads, seven repetitions).}
\label{tab:retention}
\begin{tabular}{rrrrr}
\toprule
Hot set & Normal, cold 256 & Persist, cold 256 & Persist benefit & Cold 0 baseline \\
(MiB) & (cycles) & (cycles) & (cycles) & (cycles) \\
\midrule
16 & 324.73 & 279.37 &  45.36 & 279.39 \\
32 & 370.00 & 279.37 &  90.64 & 279.38 \\
48 & 418.69 & 282.71 & 135.98 & 282.70 \\
64 & 465.74 & 284.36 & 181.38 & 284.36 \\
72 & 488.95 & 333.81 & 155.14 & 284.92 \\
80 & 512.14 & 512.08 &   0.07 & 285.37 \\
88 & 536.07 & 535.92 &   0.15 & 287.67 \\
\bottomrule
\end{tabular}
\end{table}

\section{Sustained High-Utilization Controls}
\label{sec:sustained}

\begin{figure}[H]
\centering
\includegraphics[width=0.92\linewidth]{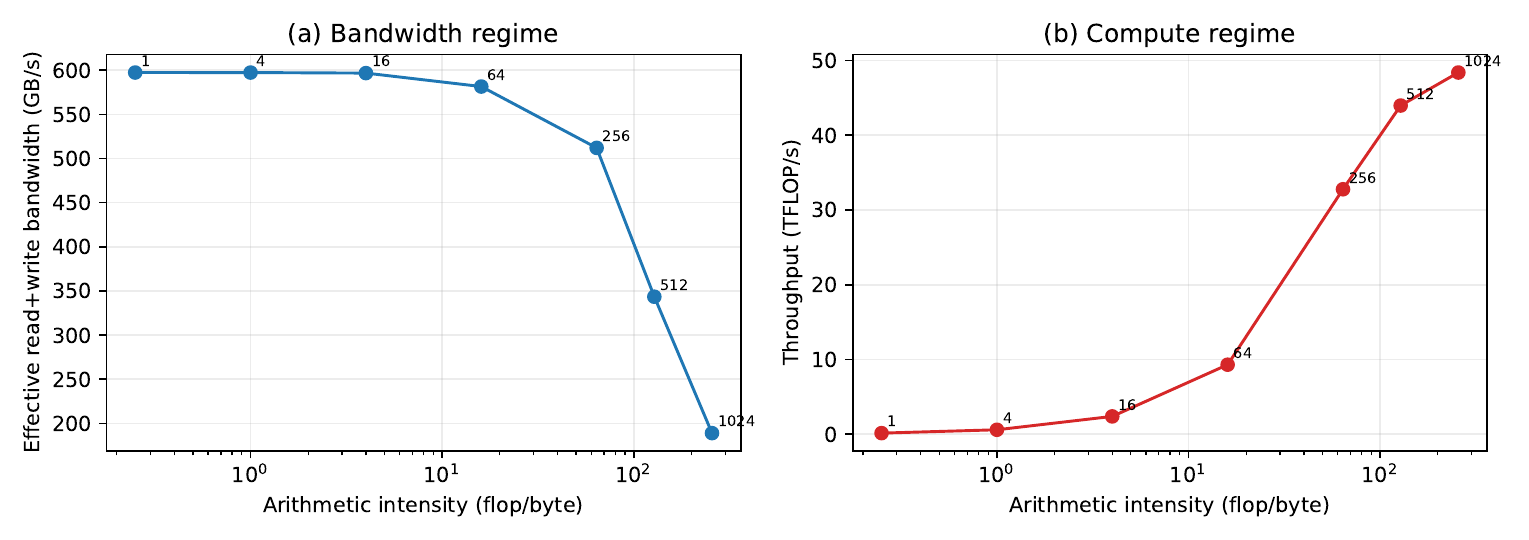}
\caption{Sustained \SI{40}{GiB} operating-regime sweep. Low arithmetic intensity
is bandwidth-limited near \SI{597}{GB/s}; high intensity becomes power-limited
near \SI{300}{W} and reaches \SI{48.38}{TFLOP/s}.}
\label{fig:roofline}
\end{figure}

Latency microbenchmarks underutilize the GPU. To record the operating envelope of
the same instance, a separate \SI{40}{GiB} read-modify-write kernel reads and
writes one \texttt{float4} per element while varying fused multiply-add
iterations from 1 to 1024; the working set occupies 41{,}395 of the
46{,}068~MiB reported by \texttt{nvidia-smi}. Figure~\ref{fig:roofline} traces
the transition from memory-bound to power-bound execution: low arithmetic
intensity sustains a median \SI{597.4}{GB/s} read+write rate at 100\% reported
utilization, and high intensity reaches \SI{48.38}{TFLOP/s} as board power meets
the \SI{300}{W} limit. The instance reaches both its memory-bandwidth and
board-power envelopes in the same session as the latency study.

\section{Related Work and Discussion}

GPU memory hierarchies have been reverse-engineered by microbenchmarking for
several generations, establishing cache sizes, line sizes, latencies, and
associativities for Volta, Ampere, Hopper, and earlier parts
\cite{mei2017dissecting,jia2018volta,luo2025hopper,zhang2020delta,vanstigt2022isolating}.
These studies report a representative L2-hit latency per device. Our result is
that on the L40 a single representative number discards a 52\% effect: the hit
latency is a function of physical SM placement.

The closest work is the real-GPU NoC characterization of
Jin~et~al.~\cite{jin2024noc}, which measured core-to-partition latency on V100,
A100, and H100 and found non-uniformity driven by physical placement, with a
discrete step across the A100/H100 dual-partition boundary and an approximately
uniform per-core average. The L40 differs on both counts. It has no discrete
partition step---the per-SM averages form a continuous gradient---yet its per-SM
average is non-uniform by 23\%, which the multi-partition parts did not show. The
two-fold symmetry at period 72 indicates a symmetric two-cluster GPC arrangement
on the single AD102 die, and the \SI{512}{B} slice-interleave period and 57/40
cycle SM/slice terms give a compact characterization of the fabric without access
to performance counters. Together these extend real-GPU NoC measurement to the
Ada generation and to a single-partition data-center GPU that the prior study did
not include.

The practical consequences are twofold. For performance, a latency-bound,
reuse-heavy kernel pinned to a near SM sees its L2 hits roughly 57 cycles faster
than the same kernel on a far SM; Section~\ref{sec:placement} turns this into a
measured 11\% makespan reduction for NUCA-aware work placement, and performance
models that assume one L2 latency mispredict by up to 52\% on latency-bound
phases. The effect is not confined to the L40: Section~\ref{sec:crossdev} finds
the same structure on a Blackwell RTX 5090, with a slower absolute L2 and a
weaker two-fold symmetry. For security, the placement signal is a self-localization and fingerprinting
primitive: Section~\ref{sec:sidechannel} shows it leaks 6--7 bits about a kernel's
own physical scheduling (exact SM at 99.2\%), and Section~\ref{sec:samemodel}
shows two identical-model L40s are separable at 100\% from their per-SM signature.
This extends practical GPU side-channel concerns
\cite{naghibijouybari2018rendered,dutta2022spy} to placement and device
inference on Ada, while remaining a stepping stone---not a complete exploit---as
it reads the measuring kernel's own placement and extracts no victim data.

\section{Limitations}

The study covers two physical L40s and one RTX 5090, with their own
driver/runtime stacks and host platforms. Two L40s are enough to demonstrate
same-model fingerprinting but not to characterize its distribution across a
population of dies, and the cross-architecture comparison rests on a single 5090.
The security capability is scoped to self-localization and device
fingerprinting: a kernel reads its own placement and device identity. Inferring
the placement of a separate victim kernel, or extracting its data, would require
a contention or interference channel we do not measure; the primitive is a
stepping stone for de-anonymization or co-residency detection, not a complete
exploit. The container does not permit privileged performance counters, so we do
not report the measured slice-to-controller mapping,
associativity, or internal routes; the topology is inferred from timing,
its two-fold symmetry, the slice-interleave period, and cross-pattern agreement.
The mapping of \texttt{\%smid} to the physical die floorplan is consistent with
the AD102 GPC count but is not independently confirmed by a die photograph. The
latency probe uses one in-flight dependent load to isolate the fabric path;
real kernels issue independent and coalesced requests, use L1 and shared memory,
and overlap memory with computation, so the absolute cycle counts are not
application throughput. The bandwidth uniformity reported for other GPUs
\cite{jin2024noc} is not measured here; our throughput data are aggregate
controls, not per-SM.

\section{Reproducibility and Artifact}

The artifact is organized for ancillary-file submission. The TeX source, BibTeX
file, and figures remain in the submission root. Code, configurations, raw CSV
results, manifests, and regeneration scripts are placed under \texttt{anc/}. Raw
observations are CSV or JSON and compress well below the \SI{50}{MB} ancillary
budget. Every randomized chain has an explicit 64-bit seed; each run records
command lines, source SHA-256 hashes, host information, CUDA/NVIDIA versions, and
before/after telemetry. The per-SM map regenerates from
\texttt{topology\_slice\_sweep.csv} and \texttt{topology\_chain\_uniform.csv} via
\texttt{analyze\_topology.py}. The placement oracle is built from the labeled
fingerprint shots (\texttt{topology\_fp\_a*.csv.zst}) by
\texttt{train\_oracle.py}; the trained model, its probe-bank metadata, and a
held-out fingerprint sample are included so \texttt{predict\_sm.py} runs the
oracle offline without a GPU. The cross-architecture comparison
(\texttt{analyze\_cross\_device.py}), the same-model fingerprinting
(\texttt{analyze\_samemodel.py}, from the second L40's runs), the
sustained-load stability runs (\texttt{run\_stability.py},
\texttt{analyze\_stability.py}, with the \texttt{topology\_stability} kernel), and
the placement campaign (\texttt{run\_placement.py}, \texttt{analyze\_placement.py},
with the \texttt{placement\_sched} kernel) all regenerate from their archived
runs. Reproducing the GPU measurements requires an NVIDIA L40 or a compatible
device (the cross-device run additionally uses a Blackwell RTX 5090), and the
scripts record changed device properties in a new manifest rather than
overwriting archived results.

\section{Conclusion}

On the tested NVIDIA L40 the L2-hit latency is not a single number but a function
of physical SM placement, spanning \num{222.5}--\num{339.2} cycles (52\%) across
the 142 SMs. An additive model with a \num{57.2}-cycle SM term and a
\num{39.5}-cycle slice term explains 87\% of the variation, the SM term is
two-fold symmetric at period 72 in line with the AD102 GPC layout, and the
effect is confirmed physical by independent access patterns ($r=1.000$) and a
sub-\num{0.01}-cycle noise floor. The L40 lacks the discrete inter-partition step
of A100/H100 but, unlike them, has a non-uniform per-SM average. The map is a
stable per-die signature: it is unchanged after an hour at full load, two
identical-model L40s are separable at 100\% from it despite near-identical mean
latency, and it lets a kernel self-localize to its own SM at 99.2\%---a
fingerprinting primitive, not a victim attack. The same structure appears on a
Blackwell RTX 5090, so the finding is not L40-specific, and distributing
latency-bound work by the per-SM map cuts makespan by up to 11\%---a concrete use
of the topology, bounded to the latency-bound regime. The single-thread residency
results---an effective-capacity boundary near \SI{96}{MiB}, line-tag scaling,
a prefetch-modifier null result, and a \SI{66}{MiB} persisting boundary---hold as
controls, but each is an average over the non-uniform latency this paper maps.

\bibliographystyle{plain}
\bibliography{refs}

\end{document}